# A Biomechatronic Approach to Evaluating the Security of Wearable Devices in the Internet of Medical Things

Yas Vaseghi[a], Behnaz Behara[b], and Mehdi Delrobaei[c,*]

[a] Department of Systems and Control, Faculty of Electrical Engineering, K. N. Toosi University of Technology, Tehran, Iran.
[b] Department of Biomedical Engineering, Faculty of Electrical Engineering, K. N. Toosi University of Technology, Tehran, Iran.
[c] Department of Mechatronics, Faculty of Electrical Engineering, K. N. Toosi University of Technology, Tehran, Iran.

**ARTICLE INFO**

This work was supported by K. N. Toosi University of Technology's Chancellor's Visionary Grant.

**ABSTRACT**

The Internet of Medical Things (IoMT) has the potential to revolutionize healthcare by reducing human error and improving patient health. For instance, wearable smart infusion pumps can accurately administer medication and integrate with electronic health records. These pumps can alert healthcare professionals or remote servers when an operation fails, preventing distressing incidents. However, as the number of connected medical devices increases, so does the risk of cyber threats. Wearable medication devices based on IoT attached to patients' bodies are particularly vulnerable to significant cyber threats. Since they are connected to the internet, these devices can be exposed to potential harm, which can disrupt or degrade device performance and harm patients. Therefore, it is crucial to establish secure data authentication for internet-connected medical devices to ensure patient safety and well-being. It is also important to note that the wearability option of such devices might downgrade the computational resources, making them more susceptible to security risks. We propose implementing a security approach for a wearable infusion pump to mitigate cyber threats. We evaluated the proposed architecture with 20, 50, and 100 users for 10 minutes and repeated the evaluation 10 times with two infusion settings, each repeated five times. The desired volumes and rates for the two settings were 2 ml and 4 ml/hr and 5 ml and 5 ml/hr, respectively. The maximum error in infusion rate was measured to be 2.5%. We discuss the practical challenges of implementing such a security-enabled device and suggest initial solutions.

*Corresponding address: Department of Mechatronics, Faculty of Electrical Engineering, K. N. Toosi University of Technology, Tehran, Iran.
E-mail address: delrobaei@kntu.ac.ir

## 1. Introduction

According to statistics from the National Center for Chronic Disease Prevention and Health Promotion (NCCDPHP), 60% of adults in the US are affected by at least one chronic disease, and 40% have two or more [1]. Patients with chronic diseases often require continuous symptom control and medication administration [2]. However, frequent visits to healthcare facilities can be challenging for many patients, disrupting their daily lives. Furthermore, exceptional circumstances like the COVID-19 pandemic pose risks to patients and their families, highlighting the need for alternative solutions. In this context, the development and implementation of remote patient monitoring and management systems are crucial.

Smart healthcare integrates platforms that link individuals, resources, and institutions via wearable technology, IoT, and mobile connectivity [3], [4]. The central goal of smart healthcare is to provide prompt medical services [5] as the incidence of chronic illnesses rises [6] and the demographic shifts towards an older population, presenting substantial obstacles for conventional medical systems. Consequently, to avoid overburdening healthcare infrastructures, in-home telehealth systems are crucial for the future of medical care [7].

Our research is focused on enhancing smart healthcare by designing automated and secure systems for managing medication delivery. Conventional methods often struggle to ensure the effectiveness of treatments [8]. These challenges are primarily due to the need to deliver the correct medication dosage precisely at the ideal interval. Consequently, there has been a move towards refining these factors [9]. This study presents the development of a smart and secure wearable infusion pump, advancing towards a reliable closed-loop system for managing chronic conditions such as Parkinson's disease [8]. Smart wearable infusion pumps ensure accurate medication dosage and adjust treatment based on real-time monitoring of vital signs, customizing care to the individual needs of each patient [3].

This study aims to develop a secure authentication mechanism using tokens and propose a smart infusion pump. A user-friendly web application interface has been created to allow authorized patients and physicians to access patient data. Each patient's physician can define drug volume and rate limits in their profile, making drug delivery personalized.

The server's performance was evaluated across three user groups with varying numbers of users. The evaluation focused on two key performance measures: throughput and average response time. The system was comprehensively evaluated by measuring infusion volume and rate and calculating the associated errors.

This paper is organized as follows: Section II provides an overview of prior research in the same field. Section III outlines the methodology, system architecture, and experimental setup. Section IV showcases the performance evaluation results. Finally, Section V provides the study's conclusion.

## 2. Related Work

Zheng et al. [10] proposed an authentication system known as the "Finger-to-Heart" (F2H) system to ensure the security of wireless implantable medical devices (IMDs). This system uses a patient's fingerprint for authentication and access permission to the IMDs. The focus is on minimizing resource use in their biometric security approach. In this scheme, the IMD does not need to capture and process biometric data during every access attempt. Moreover, they proposed an improved fingerprint authentication algorithm. proposed.

Kulaç [11] developed an external protective system that uses proxy technology to improve the safety of wireless implantable medical devices (IMDs) lacking inherent security. This system employs full-duplex technology to guarantee secure communication for these devices. Furthermore, the addition of sensors to the protective jacket enhances physical layer security.

Jamroz et al. [12] introduced an authentication approach for Internet of Medical Things (IoMT) devices utilizing hyperelliptic curves and incorporating dual signatures. The reduced key size associated with hyperelliptic curves contributes to the method's efficiency. Additionally, an analysis of security validation showed that the proposed method is resilient against various attacks.

Yu et al. [13] developed a smart insulin patch that forms part of a closed-loop system designed to administer insulin as blood glucose levels rise. The study focused on issues such as fast responsiveness, simple administration, and biocompatibility.

Wu et al. [14] introduced a self-powered wearable transdermal drug administration system featuring a hydrogel skin patch equipped with adjacent electrodes for medication dispensation. The system harnessed energy from biomechanical movements to support a closed-loop drug delivery mechanism. Designed for non-invasive infusion, the patch employs iontophoresis, a technique that uses an electric current to transport charged particles into the skin.

Kamarei et al. [15] presented an approach to enhance the security and efficiency of IoT-based healthcare systems. Their study addresses the challenges posed by congestion in wireless body area networks (WBANs), which can significantly affect the performance of medical IoT systems. To mitigate the challenges caused by congestion, the authors proposed an efficient algorithm that dynamically partitions WBAN nodes within an IoT system to balance traffic load and prevent congestion. Each partition is managed by an Access Point (AP) that adjusts its coverage range based on the traffic load to distribute the load evenly among APs and alleviate congestion. Additionally, the study employed fuzzy logic to manage the computational demands of the method, making it suitable for resource-limited IoT devices common in healthcare settings.

Rathore et al. [16] proposed a multi-layer security scheme for implantable medical devices (IMDs) in the context of the Internet of Medical Things, which included an electrocardiogram (ECG) authentication scheme using Legendre approximation and a multi-layer perceptron (MLP) model. The proposed scheme aimed to provide three levels of security for data, network, and application layers in IMDs.

The authentication scheme using ECG signals coupled with the MLP model achieved up to 99.99% testing accuracy in identifying authorized personnel.

Masud et al. [17] introduced a lightweight and physically secure mutual authentication and secret key establishment protocol for IoMT in COVID-19 patient care. Their system uses physical unclonable functions (PUFs) to verify the legitimacy of the doctor and sensor node before establishing a session key. The protocol protects the sensor nodes from tampering, cloning, and side-channel attacks, making it suitable for deployment in unattended and hostile environments. The protocol was tested using the AVISPA tool, which confirmed its safety against prominent attacks, including replay, and man-in-the-middle.

## 3. Method

### 3.1. System Architecture

A popular architecture for healthcare IoT systems generally consists of three layers: the sensor or things layer, the communication layer, and the server or processing layer [18-21]. Our model's system architecture also consists of three layers: the application, network, and data storage layers. The physician and an IoT device connected to the patient are included in the application layer. An application programming interface (API) was designed as the network layer to act as an intermediary between the application and data storage layers. Finally, the data storage layer represents the database that stores all the required information for the entire process. Fig. 1 provides an overview of the proposed system's general architecture, while Fig. 2 shows a more detailed representation of the implemented model and authorization process.

1) Application layer: In our proposed system architecture, the application layer is the uppermost layer, enabling interactions between the physician and the IoT device. Specifically, a wearable infusion pump constitutes the IoT device within this architecture. This infusion pump can connect to the internet, thereby facilitating data exchange with the data-storage layer via the designed network layer.

2) Network layer: The network layer incorporates two APIs and is tasked with facilitating connectivity between the application player and the data storage layer. It enables the IoT device and the physician to transmit and receive data to and from the data storage layer. The communication protocol employed to link these layers is the Hypertext Transfer Protocol (HTTP), a widely recognized method for data transmission across the Internet. HTTP operates on a request-response model wherein the client (in this case, the application layer) issues an HTTP request to a server. The server processes this request and issues an HTTP response containing the client's requested data. The client and the server communicate via sequential request-response exchanges. The next section will explore the designed APIs and their specific functionalities.

3) Data storage layer: In the developed application, accounts are established for both physicians and patients. Physicians are granted access to real-time vital data of patients. They are authorized to either approve or reject modifications and adjustments to drug volume or rate suggested by the algorithm. Moreover, all infusion activities are recorded in a database, thereby enabling physicians to review the patient's infusion history for a thorough analysis of their treatment. Additionally, each time an infusion pump linked to an authorized account sends a request, the requested information is retrieved from the data storage layer, cached, and subsequently transmitted back to the pump.

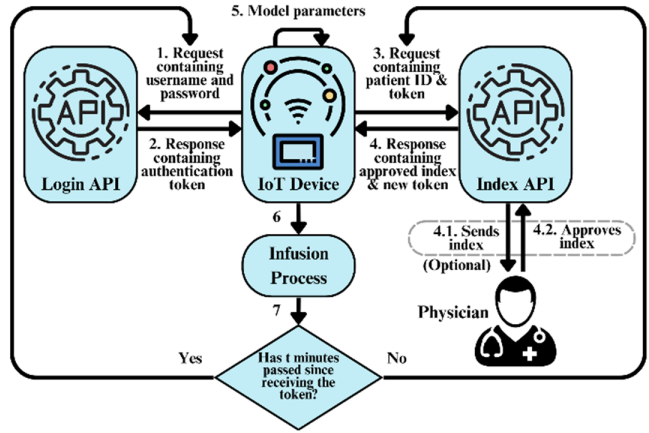

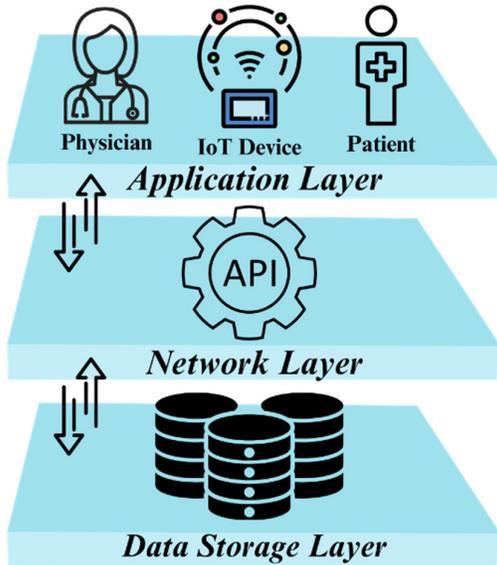

Figure 1. The proposed system's overall structure. The IoT device and physician comprise the application layer. The database forms the data storage layer, while APIs comprise the network layer.

Figure 2. The detailed representation of the system, which incorporates two APIs. The authorization process utilizes a keyless algorithm that uses token-based security.

### 3.2. Experimental Setup

This section presents the design methodology, and the hardware components employed in the fabrication of the syringe infusion pump, then discusses the integration of software that controls its functioning and ensures a secure link to the data storage layer. The infusion pump, designed to be compact and wearable, can be worn around the waist or transported in a bag. The performance of the infusion pump was evaluated in accordance with the framework outlined in Fig. 3.

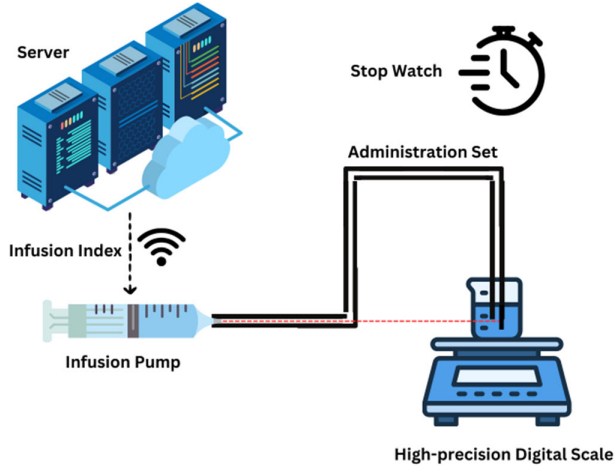

Figure 3. The diagram of the experimental setup. The infusion index is sent to the pump through the Internet, and the pump's performance is evaluated by tracking the weight of the medication administered at each instance.

1) Hardware: Syringe infusion pumps are electromechanical systems that exhibit an accuracy spectrum from approximately ±2% to ±5%. Notably, these pumps are capable of sustaining both precision and accuracy at low flow rates [22]. Consequently, in this study, the syringe mechanism was selected as the method of infusion for the pump.

For precise control of the syringe plunger of the infusion pump, an Actuonix P8 miniature linear stepper motor was selected due to its precise 0.0018-millimeter full-step size. This motor offers precision, a lightweight design of only 28 grams, and is well-suited for low-voltage wearable devices. Operating at a maximum input voltage of 4.2 volts and a current per phase of 256 milliamperes, the motor meets the energy efficiency criteria required for wearable technology. Furthermore, its 165:1 gear ratio enhances its ability to handle viscous fluids effectively. Integrated with a Pololu DRV8834 low-voltage stepper motor driver, which includes features such as thermal and current protections, adjustable current control, and a voltage range of 2.5-10.8 volts, the motor's operations are managed by an ESP32-DevKitC V4, which handles connectivity, the conversion of infusion volume into motor steps, and the management of the user interface. The device is powered by a 3.7 V 3400 mAh lithium-ion rechargeable battery and a T6845-C power bank module. This setup ensures continuous operation, even at low battery levels, by allowing power to be drawn from external sources via a micro-USB port. The specifications of the fabricated infusion pump are detailed in Table I.

TABLE I. SPECIFICATIONS OF THE FABRICATED INFUSION PUMP

| Size (cm) | Weight (g) | Power Supply | Infusion Rate (mL/h) | Error |
|---|---|---|---|---|
| 20×5×4.8 cm | 316 | lithium-ion rechargeable battery | 0.1 to 200 | Less Than 3% |

2) Software: Our proposed IoT device emphasizes both security and precision. To safeguard against unauthorized data access via the infusion pump, we have implemented time-expiring, single-use tokens as a robust security mechanism. Subsequent sections will explore the authentication process in detail. Fig. 4 illustrates the developed web application.

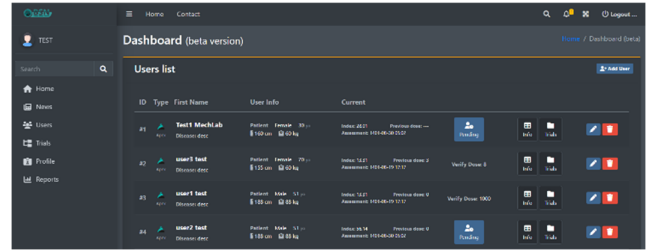

Figure 4. A screenshot of the developed web application.

a) Authorization: As mentioned, our IoT device incorporates two APIs: The login API and the index API. The system utilizes a keyless security mechanism through the use of token-based authentication. Upon entry of the user's designated username and password into the pump's interface, the microcontroller leverages these credentials as parameters in an HTTP POST request sent to the login API for authentication. In response, if the credentials match those stored on the server, the application responds back to the microcontroller. This response includes four parameters: The user's first name, last name, affiliated institution, and a single-use, time-expiring token. This token serves as the authorization credential enabling access to resources managed by the index API.

Upon receiving the response, the microcontroller extracts the provided token. This token is then utilized as a parameter in a subsequent HTTP POST request directed to the index API. This request also uses the patient ID as a parameter. Successful authentication with the token prompts the index API to return a response containing two key parameters: an infusion index and a new single-user, time-limited token. The microcontroller processes these parameters, initiating the infusion based on the specified index and retaining the new token for future communication with the index API once the current infusion completes and a subsequent index is required. The token issued by the index API is time-sensitive and expires after a predetermined duration. Should a request be made with an expired token resulting in failed authentication, the microcontroller will initiate a new request to the login API to undergo the authentication process again. Fig. 5 and 6 illustrate the HTTP message formats used by the login and index APIs, respectively.

Moreover, each account in the system is linked to a particular patient device via the device's Media Access Control (MAC) address. As a unique identifier for network interfaces, the MAC address serves as a critical element in security and verification protocols, restricting account access to only pre-registered devices within the system. This precaution significantly fortifies the security of patient data by preventing unauthorized access and upholds the integrity and reliability of the data transmitted between the device and the system. Therefore, access is limited to the devices officially recorded under the respective user's account, ensuring controller and secure interaction.

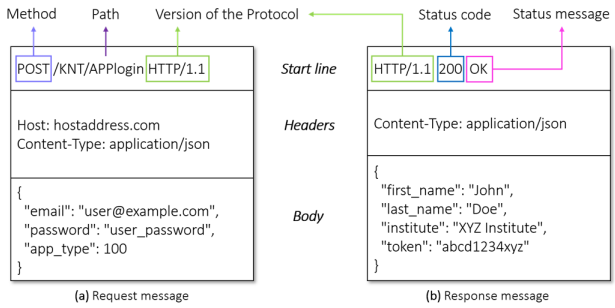

Figure 5. A representation of the request and response messages (login API).

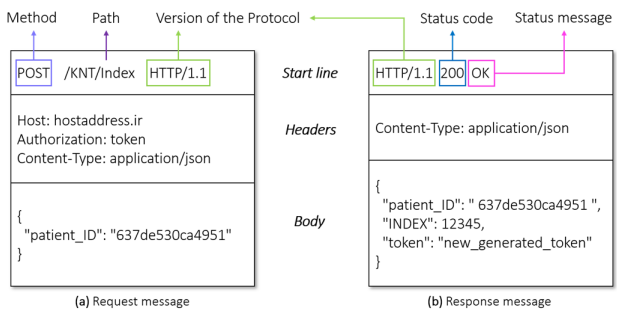

Figure 6. A representation of the request and response messages (index API).

b) Polling: The decision-making algorithm, integral to the operation of the infusion system, may require adjustments to the infusion index even during active infusion sessions. Such modifications could stem from several factors, including changes in the patient's physiological condition. As a result, the design of the infusion pump includes an autonomous feature that checks its connection to the APIs at set intervals. This functionality ensures that the pump remains consistently synchronized with the most current data, facilitating timely and accurate adjustments to the infusion process.

During each connectivity verification, the pump retrieves the current infusion index anew. It is critical to emphasize that any deviations between the newly acquired data and the parameters of the current infusion prompt the system to activate a protocol for adjusting the delivery parameters. New commands are then formulated and transmitted to the stepper motor, which is responsible for the accurate modulation of the volume and rate of the infusion. This adaptive method guarantees that the infusion process continuously adapts to reflect the latest therapeutic directives, thereby optimizing patient treatment and enhancing the robustness of safety protocols.

Thus, the infusion pump effectively collaborates with the algorithm, adjusting in real-time to ensure that the patients receive the most relevant and suitable medication dosage. This process is carried out proficiently and promptly, thereby reducing the potential risks or delays associated with the delivery of critical care.

## 4. PERFORMANCE EVALUATION

### 4.1. Server Capability Evaluation

Following the proposed architecture's design and implementation, we assessed the system's performance. The evaluation metrics employed to gauge the efficacy of the proposed system include throughput and response time, described below:

- Throughput: This metric refers to the number of requests sent to the server per second.

- Response time: This metric measures the duration taken for recording patients' data in the system, updating and receiving information between the physician and the patient.

The proposed architecture's performance was evaluated with three different user groups, including groups of 20, 50, and 100 users. The duration of the performance evaluation was considered 10 minutes. Figs. 7 and 8 show the performance evaluation for 20, 50, and 100 users.

Fig. 7 presents the average throughput observed across three distinct user groups over 600 seconds. The data from the graph suggest that a rise in system throughput accompanies an increase in the number of users. Fig. 8 illustrates the average response time for these user groups within the same temporal window. The graph demonstrates that the average response time escalates as the number of users increases.

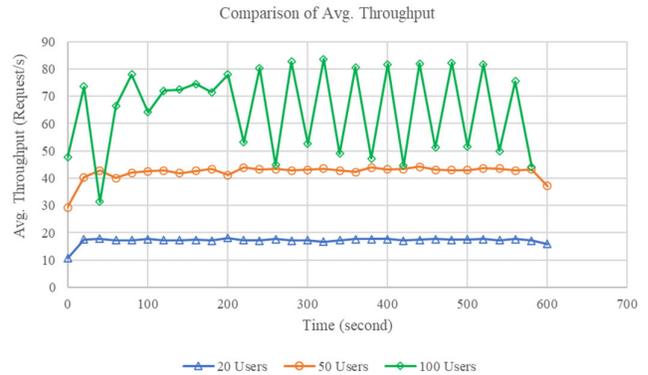

Figure 7. Comparison of average throughput between three user groups.

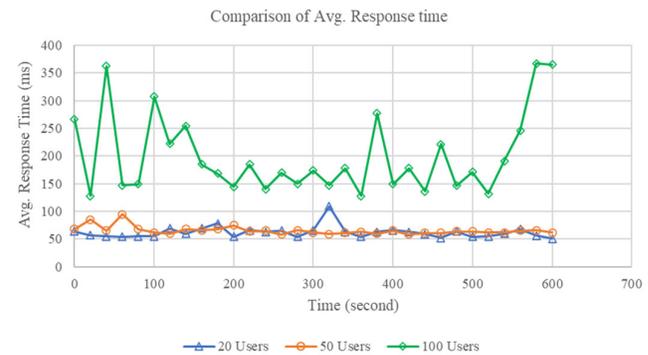

Figure 8. Comparison of average response time between three user groups.

Table II shows the total number of requests, average throughput, maximum response time, minimum response time, and average response time for three user groups.

TABLE II. DIFFERENT EVALUATION PARAMETERS FOR THREE USER GROUPS

| User group | Total requests | Avg. response time (ms) | Max. response time (ms) | Min. response time (ms) | Avg. Throughput (requests/sec) |
|---|---|---|---|---|---|
| 20 | 10454 | 62 | 1305 | 42 | 17.02 |
| 50 | 25483 | 66 | 2798 | 41 | 41.32 |
| 100 | 38875 | 184 | 16424 | 42 | 63.21 |

### 4.2 Infusion Accuracy Evaluation

We employed the setup demonstrated in Fig. 9 to evaluate the system's performance and accuracy in delivering the desired volume of medication at the desired infusion rate.

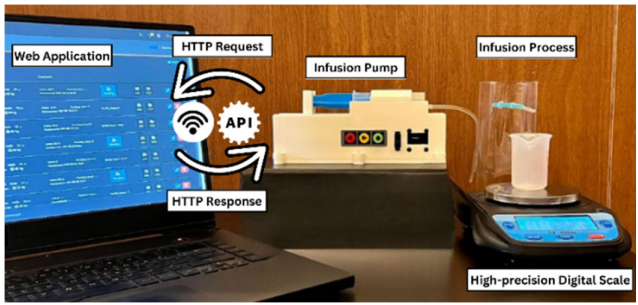

Figure 9. The experimental setup to evaluate the performance of the system.

The server sends the determined infusion volume and rate to the pump to initiate the infusion process. We employed a scale with an accuracy of 0.01 grams to measure each drop dispensed by the pump. Additionally, a stopwatch was activated at the beginning of the process to precisely record the timing of each drop as it was administered. Utilizing this data, we computed the volume of each drop infused and the following infusion rate.

This procedure was repeated 10 times across two distinct infusion settings. The first setting, which involved administering a volume of 2 milliliters and a rate of 4 milliliters per hour, was repeated five times. Similarly, the second setting, designed to deliver a desired volume of 5 milliliters and a rate of 5 milliliters per hour, was also executed five times.

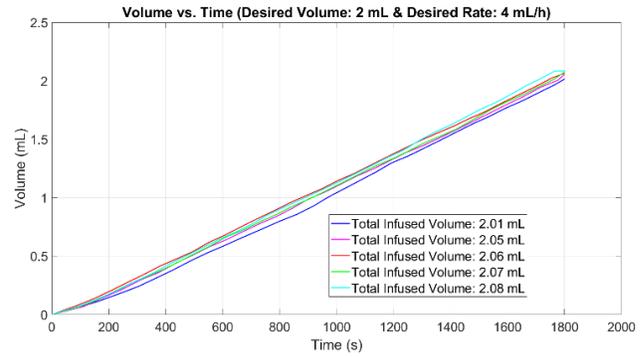

Figure 10. Infused volume over time for all five experiments of the first setting (desired volume: 2 ml, desired rate: 4 ml/h, infusion time: 1800 s).

Fig. 10 demonstrates the infused volume over time for all five experiments of the first setting.

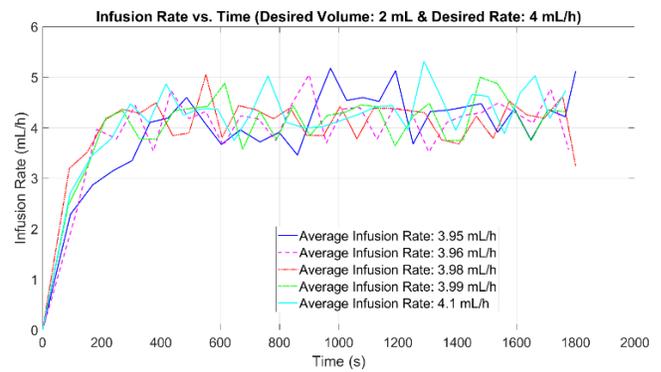

Figure 11. Infusion rate over time for all five experiments of the first setting (desired volume: 2 ml, desired rate: 4 ml/h, infusion duration: 1800 s).

Fig. 11 demonstrates the infusion rate over time for all five experiments of the first setting.

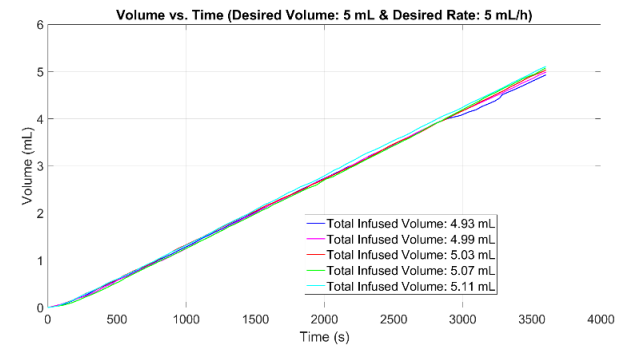

Figure 12. Infused volume over time for all five experiments of the second setting (desired volume: 5 ml, desired rate: 5 ml/h, infusion duration: 3600 s).

Fig. 12 demonstrates the infused volume over time for all five experiments of the second setting.

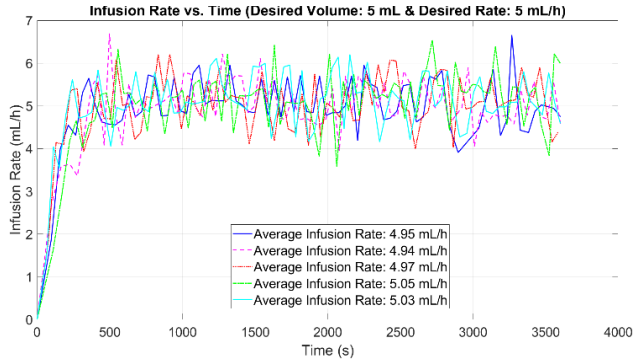

Figure 13. Infusion rate over time for all five experiments of the second setting (desired volume: 5 ml and desired rate: 5 ml/h, infusion duration: 3600 s).

Fig. 13 demonstrates the infusion rate over time for all five experiments of the second setting.

The final results of the performed experiments have been summarized in Table III.

TABLE III.  SYSTEM EVALUATION RESULTS

| Setting | Experiment | Delivered Volume (mL) | %Error in Volume | Average Infusion Rate (mL/h) | %Error in Infusion Rate |
|---|---|---|---|---|---|
| Volume = 2 mL Rate = 4 mL/h | 1 | 2.05 | 2.5% | 3.96 | 1% |
|  | 2 | 2.07 | 3.5% | 3.99 | 0.25% |
|  | 3 | 2.01 | 0.5% | 3.95 | 1.25% |
|  | 4 | 2.08 | 4% | 4.1 | 2.5% |
|  | 5 | 2.06 | 3% | 3.98 | 0.5% |
|  | Avg. %Error in Volume | | 2.7% | Avg. %Error in Infusion Rate | 1.1% |
| Volume = 5 mL Rate = 5 mL/h | 1 | 5.03 | 0.6% | 4.97 | 0.6% |
|  | 2 | 4.93 | 1.4% | 4.95 | 1% |
|  | 3 | 5.07 | 1.4% | 5.05 | 1% |
|  | 4 | 5.11 | 2.2% | 5.03 | 0.6% |
|  | 5 | 4.99 | 0.2% | 4.94 | 1.2% |
|  | Avg. %Error in Volume | | 1.16% | Avg. %Error in Infusion Rate | 0.88% |

## 5. Conclusions

This paper detailed a data authentication process aimed at enhancing the safety of an infusion pump that can be accessed and managed remotely. The principal goal of this work was to focus on patient care for chronic illnesses by implementing a secure approach for the IoT-enabled wearable infusion pumps. The system's multi-layered architecture facilitated efficient communication in the proposed configuration, while the web interface allowed authorized patients and physicians to customize drug delivery settings. Security was mainly ensured by generating unique tokens that expire immediately after the first use, and each account is tied to a specific device through its MAC address. Moreover, the infusion pump adjusts autonomously to changing therapeutic requirements to ensure prompt and safe medication administration. The system's server performance assessment indicated that it can handle multiple user groups, and the overall design was found to be reliable and practical. Expanding on this study, our team aims to develop a fully automated closed-loop monitoring and management system tailored for chronic diseases such as Parkinson's disease. This system will integrate data acquisition via motion-capture sensors, a cloud-based decision support algorithm for precise dosage determination, and the secure smart wearable infusion pump described herein. This work is intended to be a component of a larger system, where future research will concentrate on developing amd implementing the additional elements required to realize this comprehensive solution.

## 6. Acknowledgements

This work was supported by K. N. Toosi University of Technology's Chancellor's Visionary Grant.